\documentclass[ reprint, superscriptaddress, amsmath,amssymb,prl, longbibliography]{revtex4-2}

\usepackage[normalem]{ulem}
\usepackage[dvipsnames]{xcolor}

\newcommand{\ric}[1]{\textcolor{black}{#1}}

\usepackage{amsmath}
\usepackage{amsfonts}
\usepackage{pgfplots}
\DeclareUnicodeCharacter{2212}{−}
\usepgfplotslibrary{groupplots,dateplot}
\usetikzlibrary{patterns,shapes.arrows}
\usepackage{bbm}
\usepackage{cancel}
\usepackage{braket}
\usepackage{dsfont}
\usepackage{enumerate}
\usepackage{bm}
\usepackage{url}
\begin{document}
\title{Rydberg platform for non-ergodic chiral quantum dynamics}
\author{Riccardo J. Valencia-Tortora} \email[]{Corresponding author: rvalenci@uni-mainz.de}
\affiliation{Institut f\"{u}r Physik, Johannes Gutenberg-Universit\"{a}t Mainz, D-55099 Mainz, Germany}   
\author{Nicola Pancotti}
\affiliation{AWS Center for Quantum Computing, Pasadena, CA 91125, USA}
\author{Michael Fleischhauer}
\affiliation{Department of Physics and Research Center OPTIMAS,
University of Kaiserslautern-Landau, D-67663 Kaiserslautern, Germany}
\author{Hannes Bernien}
\affiliation{
Pritzker School of Molecular Engineering, University of Chicago, Chicago, Illinois 60637, USA}
\author{Jamir Marino}
\affiliation{Institut f\"{u}r Physik, Johannes Gutenberg-Universit\"{a}t Mainz, D-55099 Mainz, Germany}
\date{\today}%
\begin{abstract}
\noindent
We propose a mechanism for engineering chiral interactions in Rydberg atoms via a \textit{directional} antiblockade condition, where an atom can change its state only if an atom to its right (or left) is excited. The scalability of our scheme enables us to explore the many-body dynamics of kinetically constrained models with unidirectional character. We observe non-ergodic behavior via either scars, confinement, or localization, upon simply tuning the strength of two driving fields acting on the atoms. We discuss how our mechanism persists in the presence of classical noise and how the degree of chirality in the interactions can be tuned, opening towards the frontier of directional, strongly correlated,  quantum mechanics using neutral atoms arrays.
\end{abstract}
\maketitle

\textit{Introduction.--}
Despite being far from full fault-tolerant quantum computing~\cite{Preskill2018}, reliable analog quantum simulations are nowadays attainable across a variety of atomic, molecular optical (AMO) as well as solid-state platforms. Among the former, atomic Rydberg arrays stand out prominently due to their remarkable degree of programmability, as highlighted in various studies~\cite{Bernien2017, Scholl2021, Bluvstein2021, Browaeys2020, Labuhn2016}. This has led to groundbreaking experiments in areas such as topological order~\cite{deLsleuc2019, Semeghini2021}, engineerable quantum phase transitions~\cite{Bernien2017, Ebadi2021, Keesling2019, Scholl2021, Keesling2019}, lattice gauge theories~\cite{PhysRevX.10.021041}, and strongly correlated quantum dynamics~\cite{Bernien2017, PhysRevX.8.021070, PhysRevX.8.021069, PhysRevLett.120.180502}.
Such broad flexibility suggests opportunities \ric{for} designing quantum simulators with no direct counterpart in traditional AMO or condensed matter physics.
In this context, a challenge centers on creating systems hosting non-reciprocal processes that can differentiate, particularly in one dimension, between the flow of information in the right and left directions.
{Directionality, already when restricted to \emph{single-particle} processes, has proven useful in various tasks, such as mitigating back-action effects~\cite{Kamal2011}, realizing chiral transport~\cite{PhysRevX.8.041031}, aiding the preparation of nontrivial topological states~\cite{Lodahl2017,Clerk_2022,PhysRevA.91.042116,PhysRevLett.126.043602}, and realizing unconventional phases of matter~\cite{Fruchart2021,Garrahan2018}. Thus, combining directional \textit{interactions} with the high control achieved in Rydberg platforms would pave the way for entering the realm of chiral \textit{strongly correlated} phenomena as \ric{an} uncharted frontier of quantum information processing~\cite{wei2022towards}.}

\begin{figure}[t!]
\centering
\includegraphics[width=\linewidth]{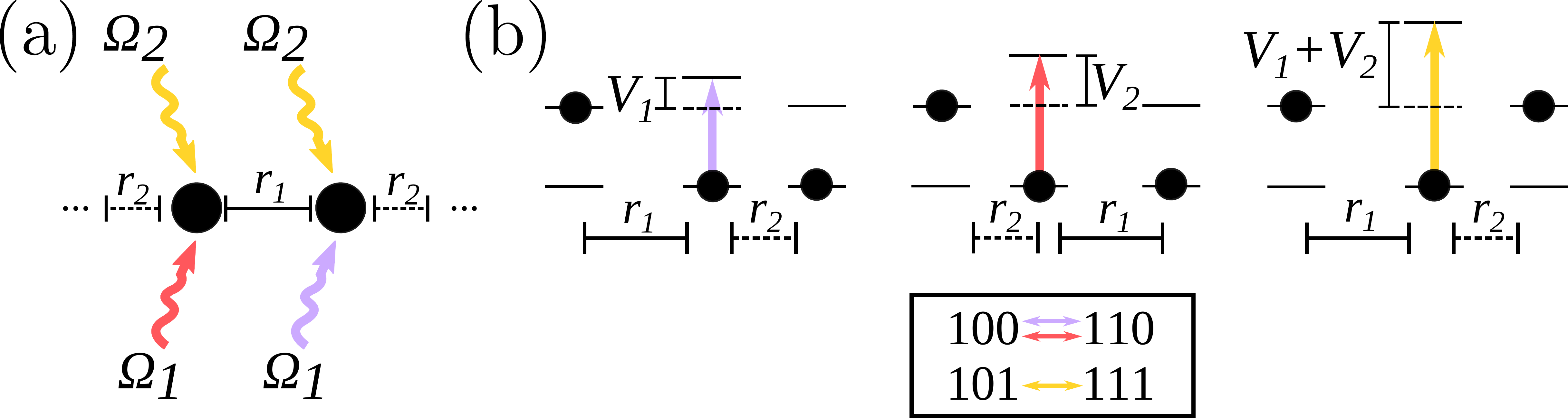}\\
\includegraphics[width=\linewidth]{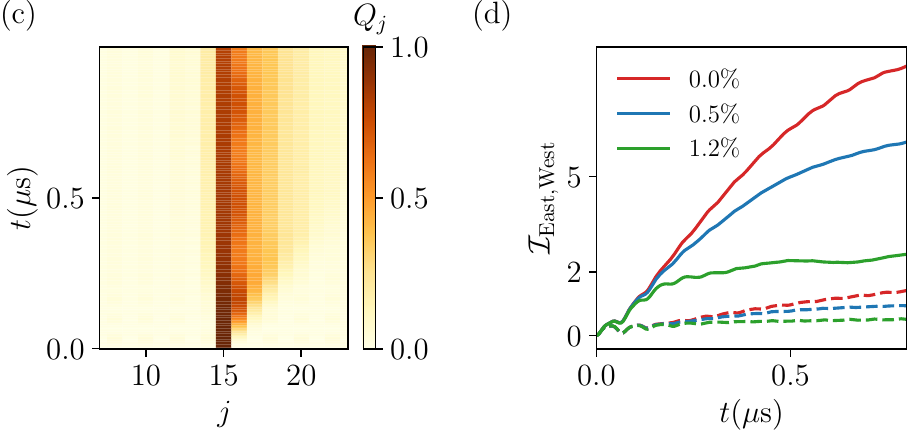}
\caption{(a): An array of Rydberg atoms in a staggered configuration of drive fields ({each color refers to a different frequency}) and spacings $r_1$ and $r_2$, with corresponding nearest-neighbor interactions $V_1$ and $V_2$. (b): Scheme of the most resonant processes (see box) which happen exclusively at the right of excited atoms due to the \textit{directional} antiblockade. (c): Dynamics of the \ric{excitation} profile seeding a single excitation \ric{for experimentally feasible parameters,} including thermal disorder in the atomic positions $\tilde{\eta}_x = 0.012$ (see text). 
(d): Dynamics of the imbalances $\mathcal{I}_\text{East}$ (continuous line) and $\mathcal{I}_\text{West}$ (dashed line) starting from the same state as in (c) for different $\tilde{\eta}_x$. The imbalance $\mathcal{I}_\text{East}$ ($\mathcal{I}_\text{West}$) is \ric{given by} the sum of the \ric{excitations}  to the right (left) of the initially seeded one. Excitations propagate preferably towards `East' as desired ($\mathcal{I}_\text{East} > \mathcal{I}_\text{West}$) despite finite temperature effects.}
\label{figure_chiral_processes}
\end{figure}

In this work, we achieve this goal by presenting a blueprint for Rydberg atomic arrays featuring chiral interactions that aren't symmetric when neighboring atoms are exchanged.  Specifically, we consider a one-dimensional array with a staggered configuration of atomic positions and drive fields (cf. Fig.~\ref{figure_chiral_processes}(a)).
In such a scenario, due to strong Van-der-Waals interactions, we can access a regime we term  \textit{directional} antiblockade, wherein an atom can change its internal state solely when an atom to its right (or left) becomes excited.
{We then show the robustness of these mechanisms to experimental imperfections, like thermal disorder in atomic positions, opening up its usage in state-of-art platforms and for simulating exotic many-body systems, such as directional kinetically constrained quantum models (KCMs).}
KCMs have attracted considerable interest due to their capability to display non-ergodic behavior despite their non-integrable and disorder-free character~\cite{PhysRevX.10.021051,PRXQuantum.3.020346,Garrahan2009,Chleboun2013,Garrahan2018,PhysRevLett.121.040603,PhysRevE.102.052132}.
In KCMs, slowdown of thermalization could occur through various mechanisms: quantum scars~\cite{Bernien2017,Turner2018,PhysRevLett.122.220603,Serbyn2021,PhysRevB.98.155134,PhysRevB.98.155134,PhysRevB.99.161101,PhysRevX.11.021021}, {realized in Rydberg arrays~\cite{Bernien2017} or with superconducting qubits~\cite{zhang2023many}}, where a few non-thermal excited states can lead to
non-relaxing dynamics; {Hilbert space shattering~\cite{PhysRevX.10.011047}, realized in ultracold atoms~\cite{PhysRevLett.130.010201,Scherg2021}}; confinement of quasi-particles induced by many-body interactions~\cite{kormos2017real,PhysRevX.10.021041,PhysRevB.102.041118,PhysRevLett.126.103002}{, observed in trapped-ions~\cite{Tan2021}}; or slow dynamics resulting from localization~\cite{PhysRevX.10.021051,PRXQuantum.3.020346}.
These mechanisms are intricately linked to the specific constraints at play, and each of them would necessitate a distinct experimental platform. Remarkably, in our Rydberg implementation, we can realize all of these mechanisms by simply adjusting the strength of the external drive fields. This versatility transforms our platform into a universal quantum simulator for non-ergodic quantum dynamics. As an additional benefit, our platform allows for the implementation of the quantum East model~\cite{PhysRevX.10.021051, PhysRevB.92.100305}, which has been absent in prior studies, albeit several Rydberg implementations have focused on related constrained models~\cite{PhysRevLett.111.215305,PhysRevA.98.021804,PhysRevA.93.040701,PhysRevA.90.011603,PhysRevA.97.011603,PhysRevA.99.060101,PhysRevLett.118.063606}.
The significance of the quantum East model lies in its distinction as one of the rare cases where an interacting system undergoes a disorder-free transition between delocalization and localization in the ground state and dynamics~\cite{PhysRevX.10.021051}, in a fashion markedly distinct from many-body localization~\cite{nandkishore2015many, RevModPhys.91.021001}.

\textit{Chiral interactions in Rydberg arrays.---}
The key ingredient to engineer a directional interaction is a staggered configuration of the atomic spacings and drive fields in a Rydberg array (cf. Fig.~\ref{figure_chiral_processes}(a)). The Hamiltonian describing such \ric{a} scenario, in the rotating frame with respect to the bare atomic transitions, can be written as
\begin{equation}
\label{eq_H_complete}
\begin{split}
&\hat{H}(t)=V_1 \sum_{j \: \text{odd}} \ric{\hat{Q}}_j \ric{\hat{Q}}_{j+1} +V_2 \sum_{j \: \text{even}} \ric{\hat{Q}}_j \ric{\hat{Q}}_{j+1} + \\
&+V_\text{NNN}\sum_{j=1}^{N-2} \ric{\hat{Q}}_j \ric{\hat{Q}}_{j+2}+\frac{1}{2}\sum_{j=1}^N \left( \ric{\Omega_j(t)} \hat{\sigma}_{j}^+ + \text{h.c.}\right),
\end{split}
\end{equation}
where $\hat{\sigma}_j^+ = |1\rangle_{j} {}_j\langle 0|$ transfers the $j$-th atom from the ground state $|0\rangle$ to the Rydberg state $|1\rangle$; $\ric{\hat{Q}}_j = |1\rangle_j {}_j\langle 1|$; $V_1$ ($V_2$) are Van-der-Waals interactions ($V(r) = C_6/r^6$) on odd (even) bonds; $V_\text{NNN}$ is the next-nearest neighbor interaction;
\ric{$\Omega_j(t)=\Omega_1 e^{-i V_{j-1,j}t} + \Omega_2 e^{-i(V_1+V_2) t}$ a classical drive field with $V_{j-1,j}$ the Van-der-Waals interaction on the $(j-1)$-th bond, and $\Omega_{1,2}$ independent Rabi frequencies}.
Throughout, we work in the regime $V_{1,2},|V_1-V_2| \gg \Omega_{1,2} \gg V_\text{NNN}$. In this regime, interactions play a crucial role in dictating the dynamics of the single atom. Specifically, two extreme scenarios can be realized: excited atoms either inhibit spin-flips of neighboring ones (blockade)~\cite{PhysRevLett.85.2208}, or facilitate them (antiblockade)~\cite{PhysRevA.76.013413}. The blockade condition occurs by setting the drive field resonant with the bare atomic transitions so that the interaction energy due to a neighboring excited atom makes it off-resonant. Instead, the antiblockade occurs when the acting drive field is detuned from the bare atomic transitions by the interaction, and thus it becomes resonant solely if a neighboring atom is excited. In translational invariant systems, each atom cannot distinguish its right neighbor from the one to its left, and so no preferable direction can appear. In our scheme instead, since $V_1 \neq V_2$, the atom can distinguish the two neighboring atoms and we can selectively make processes resonant towards one direction and off-resonance towards the other. We term this mechanism \textit{directional} antiblockade, which implies that an atom can flip only when an atom to its right (or left) is excited.
To achieve this regime, it is enough to apply \ric{the} drive field \ric{controlled by $\Omega_1$} on each atom, and so we temporarily set $\Omega_2=0$.
\ric{The drive field $\Omega_1 e^{-iV_{j-1,j}t}$ acting on site $j$ is} detuned by $V_1$ (if $j$ is even) or $V_2$ (if $j$ is odd) from the bare atomic transition, so that it
is resonant solely when the atom to its left is excited and the one to its right is not, obtaining the anticipated \textit{directional} antiblockade (see Fig.~\ref{figure_chiral_processes}(b)).
The net result is that an excitation seeded in the system triggers an avalanche of excitations solely towards `East' (see Fig.~\ref{figure_chiral_processes}(c)).
In the Supplemental Material~\cite{SupplementalMaterial}, we give further details on the experimental setup, \ric{and} an alternative scheme where the staggered configuration is imprinted on the atomic frequencies 
while the drive field is monochromatic. \\ 
\textit{Experimental realization.--} In actual experiments, the \textit{directional} antiblockade could be spoiled by finite temperature fluctuations, inhomogeneities due to the harmonic frequency trap holding the atoms, or dephasing coming from finite laser linewidth.
The first two can be taken into account
including quenched disorder in the atomic positions~\cite{ostmann2019synthetic}.
Specifically, at low enough temperature $T$,
the displacements $\delta\mathbf{r}_j$ from the ideal atomic positions are constant during a single experimental realization and distributed accordingly to a Gaussian distribution with zero average and width $\eta_\alpha =\sqrt{k_B T/(m\omega_\alpha^2)}$ along the $\alpha-$axis, with $\omega_\alpha$ the trapping frequency and $m$ the atomic mass. Instead, dephasing induced by finite linewidth of the laser can be modeled by a Lindblad master equation with jump operators $\hat{L}_j = \sqrt{\gamma}\ric{\hat{Q}}_j$, with $j \in [1,N]$. Yet, in our setup, we work in regimes where $\gamma \sim 10$ kHz is at least two orders of magnitudes smaller than the other energy scales, and therefore it can be neglected. Specifically, we will show results up to a time of $10\:\mu s$, \ric{where our approximations hold and spontaneous decay from the Rydberg state can be neglected}. We elaborate further in the conclusions and \cite{SupplementalMaterial} on the opposite limit, where dephasing is large, illustrating how our scheme readily enables us to investigate `classical' dynamics with directional character.
For concreteness, we consider ${}^{87}$Rb atoms located along the $x$-axis, at temperature $T=3\:\mu$K and optical traps with $\omega_x=\omega_y=5\times\omega_z=40\:\text{kHz}$, which give rise to an anisotropic disorder $\eta_x=\eta_y=\eta_z/5 \approx 0.1\:\mu\text{m}$.
{We consider the atomic level $70S$ as Rydberg state  which has  $C_6/(2\pi) = 864$ GHz/$(\mu$m$)^6$.} In the following, we measure disorder as the relative variation with respect to the mean distance, namely
$\tilde{\eta}_\alpha \equiv \eta_\alpha/r_1$.
We consider Rabi frequencies $\Omega_{1,2}$ in the range between $2\pi \times 1~\text{MHz}$ and $2\pi \times 5~\text{MHz}$.
\ric{For this set of parameters,
we found $V_1/(2\pi) = (15.0 \pm 1.0)$ MHz, $V_2/(2\pi)=(30.0 \pm 2.2)$ MHz, and $V_\text{NNN}/(2\pi) = (0.33 \pm 0.02)$ MHz, as a good compromise between fast dynamics,  
small impact of disorder, and $\Omega_{1,2} \ll \ric{\{V_{1,2},|V_1-V_2|\}}$. In the experiment, these interactions correspond to average spacings $r_{1}=6.2\:\mu$m  
and $r_2 = 5.4\:\mu$m, 
for which $\tilde{\eta}_x \approx 0.01$.}
Despite we show results mostly in this parameters' regime, we keep $\tilde{\eta}_\alpha$ as a free parameter to explore different experimental scenarios.
In the following, we show results averaged over $50$ realizations of disorder, for which statistical errors are $\sim 1\%$ or less.
As it can be seen in Fig.~\ref{figure_chiral_processes}(c-d), the main impact of disorder is a reduction of the propagating front,  while its directional character is not appreciably spoiled. Having shown the robustness of our scheme, we now proceed \ric{to discuss} some models immediately accessible by simply tuning the strength of the external drive fields.\\

\textit{Kinetically constrained models.---}
Before discussing the resulting dynamics of our scheme, we note that using a single drive field, an atom remains frozen when both neighbors are excited.
To enrich dynamics, we reintroduce the additional global drive field with frequency detuned by $V_1+V_2$ from the bare atomic transition, \ric{i.e. $\Omega_2 e^{-i(V_1+V_2) t}$,}  so that it allows the transition $|101\rangle \leftrightarrow |111\rangle$ (cf. Fig.~\ref{figure_chiral_processes}(b))~\cite{SupplementalMaterial}.
As it is apparent, there is still no resonant process where an atom changes its state in the absence of an excited one to its left. 
\ric{To} make it visible, we write down the following effective Hamiltonian describing the most resonant processes
without taking into account disorder in the atomic positions~(for further details see \cite{SupplementalMaterial})
\begin{equation}
\label{eq_H_eff_complete}
\begin{split}
\hat{H}&=\frac{\Omega_1}{2} \sum_j \ric{\hat{Q}}_j \ric{\hat{X}_{j+1}  \hat{P}_{j+2}} + \frac{\Omega_2}{2}\sum_j \ric{\hat{Q}}_j \ric{\hat{X}_{j+1}}\ric{\hat{Q}}_{j+2}+ \\
&+\epsilon \sum_{j} \ric{\hat{Q}}_j
+V_\text{NNN} \sum_{j}\ric{\hat{Q}}_j \ric{\hat{Q}}_{j+2},
\end{split}
\end{equation}
where \ric{$\hat{X}=|0\rangle\langle 1| + |1\rangle\langle 0|$, $\hat{P}=1-\hat{Q}$, and}
we have introduced the detuning $\epsilon$ from the perfect \textit{directional} antiblockade which, together with $V_\text{NNN}$, can spoil the perfect resonance condition and
inhibit spin-flips, as we will show later.
\begin{figure}[t!]
\centering
\qquad
\includegraphics[width=\linewidth]{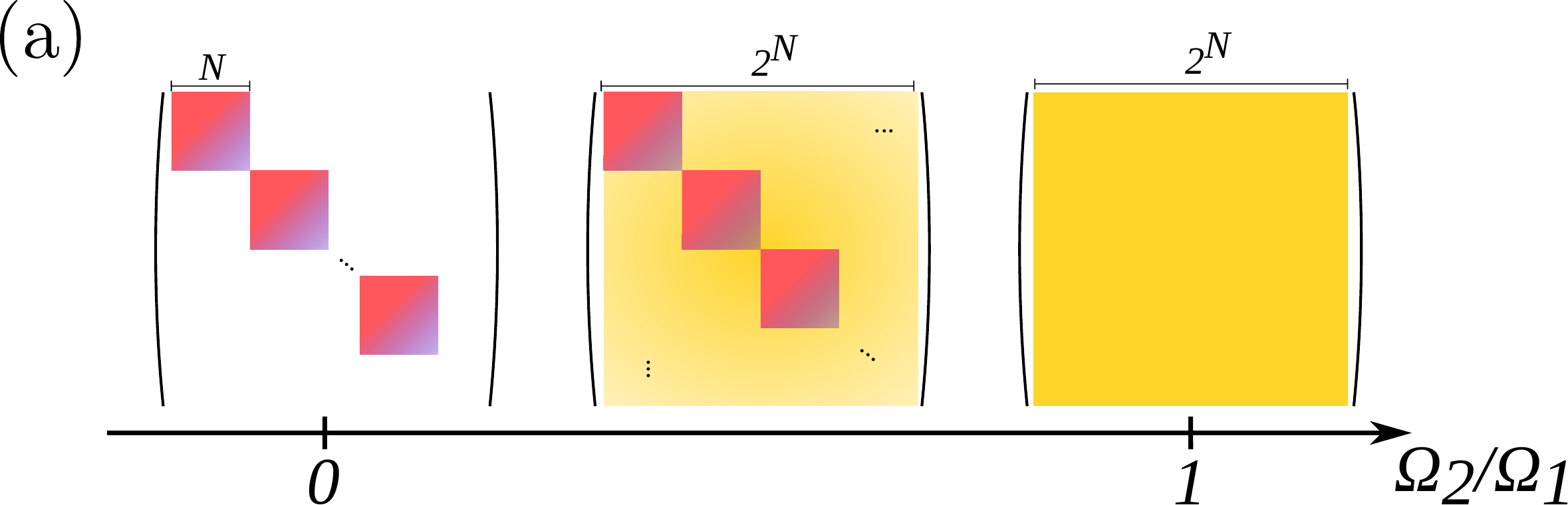}
\vspace{4pt}\\
\includegraphics[width=\linewidth]{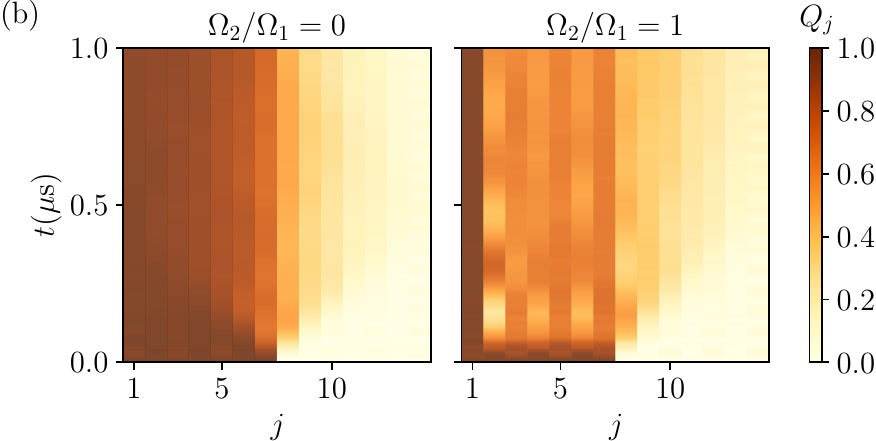}
\caption{
(a) Sketch of the accessible Hilbert space \ric{upon initializing product states}, at fixed East symmetry sector, in a system of size $N$  as a function of the Rabi frequencies $\Omega_{1,2}$. Colors indicate allowed transitions and they are in one-to-one correspondence with those used for the drive fields in Fig.~\ref{figure_chiral_processes}(a). Blank spaces indicate forbidden transitions. For $\Omega_1 \neq 0$ and $\Omega_2=0$ strings of excitations can only shrink/expand without merging/splitting, leading to Hilbert space \textit{shattering}; for $\Omega_2/\Omega_1 \neq 0$, strings achieve complete mobility, rendering the system ergodic, as any product state becomes accessible from any other.
For $\Omega_2/\Omega_1=1$ we recover the quantum East model~\cite{PhysRevX.10.021051,PhysRevB.92.100305}. (b) Dynamics of the \ric{excitation} profile starting from the product state
\ric{$\bigotimes_{j=1}^{N/2}|1\rangle \bigotimes_{j=N/2+1}^{N} |0\rangle$} in the shattered ($\Omega_2=0$) and ergodic regime ($\Omega_{1,2}\neq 0$) including
thermal disorder $\tilde{\eta}_x = 0.012$ in units of the average spacing $r_1=6.2\:\mu$m.
}
\label{figure_scheme_block_structure}
\end{figure}
\begin{figure*}[t!]
\begin{minipage}{0.48\linewidth}
 \centering
 \includegraphics[width=\linewidth]{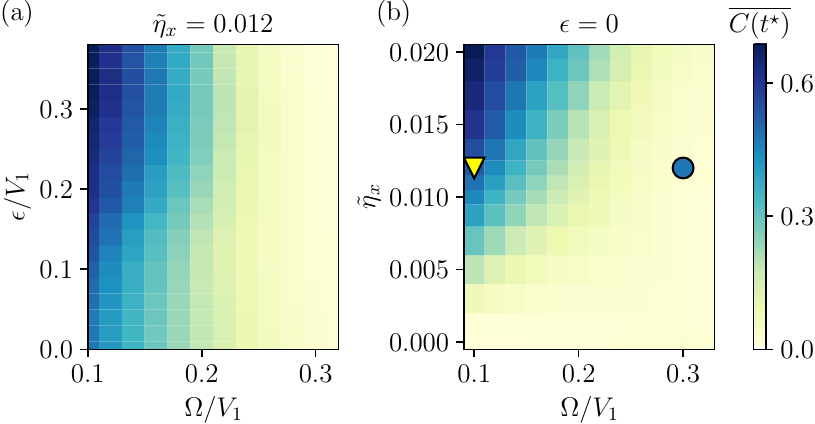}
\end{minipage} 
$\quad$
 \begin{minipage}{0.48\linewidth}
\includegraphics[width=\linewidth]{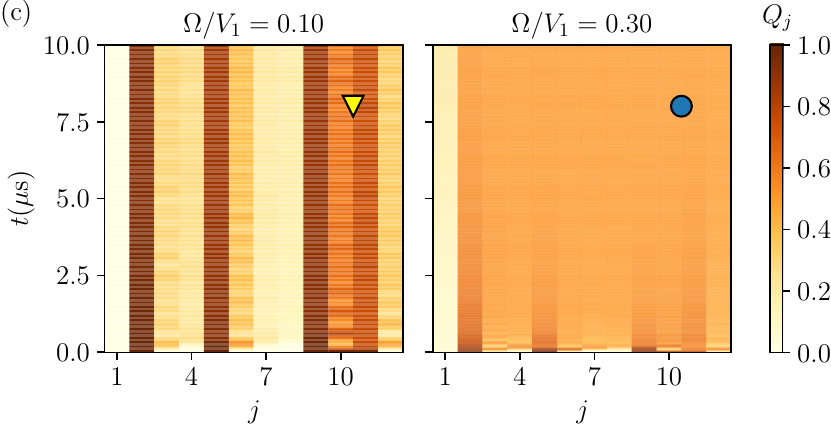}
\end{minipage}
\caption{Dynamics \ric{of the state $|010010001110\rangle$ under the full theory (cf. Eq.~\eqref{eq_H_complete})} in the quantum East model \ric{regime} ($\Omega_{1,2}=\Omega$). (a-b): Time-averaged autocorrelation function \ric{(cf. Eq.~\eqref{eq_autocorr})} $\overline{C(t^\star)}$  at time $t^\star=10\mu s$ keeping fixed either thermal disorder $\tilde{\eta}_x$ or detuning $\epsilon$, respectively.
For small $\Omega$ the memory of the initial state is kept up to long times ($\overline{C(t^\star)} > 0$), while instead for large $\Omega$ the memory of the initial state is rapidly washed out ($\overline{C(t^\star)} \approx 0$).
(c): Dynamics of the \ric{excitation} profile in the two phases (marked via symbols in (b)) at fixed thermal disorder $\tilde{\eta}_x=0.012$ in units of the average spacing $r_1=6.2\:\mu$m (see text).
}
\label{fig_autocorr_full_theory_canonical_east}
\end{figure*}
The Hamiltonian in Eq.~\eqref{eq_H_eff_complete} belongs to the family of kinetically constrained quantum East models, where dynamics are activated solely to the right of excited sites. East models are characterized by a so-called `East symmetry'~\cite{PhysRevX.10.021051}, which implies that empty regions without excitations to their left remain frozen. As a consequence, the location of the first excitation encountered starting from the left edge of the system does not change (see Fig.~\ref{figure_chiral_processes}~(c)), and the Hilbert space splits in $N$ disconnected sectors, with $N$ the system size.
Once the East symmetry sector is specified by the location of the first excitation,
we can further shape the accessible Hilbert space simply by changing the relative power of the drive fields $\Omega_1$ and $\Omega_2$ (cf. Fig.~\ref{figure_scheme_block_structure}(a)), which directly reflects in the mobility of excitations. Specifically, we will show that with a single drive field the East symmetry sector \textit{shatters} in $\mathcal{O}(e^N)$ disconnected sectors~\cite{PhysRevX.12.011050}, while when both are active it is irreducible~\cite{PhysRevX.10.021051}, meaning the dynamics connect all states
(see Fig.~\ref{figure_scheme_block_structure}). Then, we will illustrate the mechanisms by which in each regime
the onset of thermalization considerably slows down within each irreducible sector,
 exhibiting scars, confinement, and localization ({in~\cite{SupplementalMaterial} we provide further details on the dynamical features in each regime).}
In the following, we discuss these three regimes using the effective Hamiltonian in Eq.~\eqref{eq_H_eff_complete} as guidance. \ric{Then}, we show how those features survive upon simulating the full theory in Eq.~\eqref{eq_H_complete}. \\

\textit{ \ric{QXQ} model.---}
For $\Omega_1=0$ and $\Omega_2 \neq 0$ directionality is lost and a spin flip occurs solely when both neighboring atoms are excited.
This is reminiscent of the well-known PXP model, in which a spin flip occurs when both neighboring atoms are in the ground state. Indeed, the PXP model and ours share the same physics as they are connected via the transformation $\hat{U}=\prod_{j=1}^N \ric{\hat{X}}_j$, which translates to interchanging $|0\rangle \leftrightarrow |1\rangle$.
This includes Hilbert space \textit{shattering}~\cite{Moudgalya_2022} and the presence of quantum many-body scars, which slow down the onset of thermalization when initializing specific states (e.g. Néel state)~\cite{Bernien2017,PhysRevLett.122.220603}. \\

\textit{\ric{QXP model}.---}
When $\Omega_1 \neq 0$ and $\Omega_2=0$,
strings of consecutive excitations can shrink or grow but not merge or split ($|101\rangle \cancel{\leftrightarrow} |111\rangle$), effectively experiencing a repulsive interaction.
Additionally, the unidirectional character of the dynamics further reduces their mobility since the left edge of each string cannot move.
{Due to these constraints, each string of excitations is confined between its left edge and the left edge of the next one,
and no entanglement can be generated between them during dynamics.}
As a result, the Hilbert space gets shattered in $\mathcal{O}(e^N)$ disconnected sectors~\cite{PhysRevX.12.011050}.
Since each string of excitations evolves independently from the others, we will focus on the largest irreducible sector containing a single string.
We find that its dynamics {(for $\epsilon=0$)} are described by the Hamiltonian~\cite{SupplementalMaterial}
\begin{equation}
\label{eq_Heff_integrable}
\hat{H} = \frac{\Omega_1}{2} \sum_{k=1}^{N-1}\left( |\bm{k}\rangle \langle \bm{k+1}| + \text{H.c.}\right) + V_\text{NNN}\sum_{k=2}^N (k-2) |\bm{k}\rangle \langle \bm{k}|,
\end{equation}
with $|\bm{k}\rangle \equiv |1\rangle^k |0\rangle^{N-k}$, where the first term controls the change of the string length, while the second the potential energy proportional to its length.
This Hamiltonian is integrable and has been derived in similar scenarios~\cite{PhysRevLett.126.103002,PhysRevB.107.024306,PhysRev.117.432,RevModPhys.34.645}. Direct inspection shows that Eq.~\eqref{eq_Heff_integrable} is the well-known Hamiltonian of an electron in a lattice subjected to a constant electric field~\cite{PhysRev.117.432,RevModPhys.34.645}. Dynamics display Stark localization~\cite{PhysRev.117.432}, which leads to real-time Bloch oscillations
of period $T_\text{Bloch}=2\pi \Omega_1/V_\text{NNN}$ and size $\ell_\text{Bloch} \sim \Omega_1 / V_\text{NNN}$ originating from the rightmost edge of the string. Hence, excitations and quantum correlations are confined, preventing thermalization, closely resembling confinement in non-integrable systems~\cite{kormos2017real,PhysRevB.102.041118,PhysRevX.10.021041,PhysRevLett.122.150601}.   
Upon introducing disorder,
Bloch oscillations are still visible (see Fig.~\ref{figure_scheme_block_structure}(b) \ric{and \cite{SupplementalMaterial}}) although partly suppressed.

\textit{Quantum East model.---}
When both drive fields are active ($\Omega_{1,2} \neq 0$), strings of excitations gain full mobility since they can shrink, grow, merge, and split (cf. Fig.~\ref{figure_scheme_block_structure}(b)). Thus,
the accessible Hilbert space does not shatter and any product state can be dynamically reached by any other at fixed East symmetry sector~\cite{PhysRevX.10.021051}.
Nonetheless, it is still possible to observe an extreme slowdown of thermalization. This can be immediately seen by setting $\Omega_2/\Omega_1=1$, for which Eq.~\eqref{eq_H_eff_complete} reduces to the quantum East model investigated in Refs.~\cite{PhysRevX.10.021051,PhysRevB.92.100305} apart from additional density-density interactions, which do not alter the qualitative picture. \ric{This} model has been shown to display a dynamical transition separating a fast and slow thermalizing phase~\cite{PhysRevX.10.021051,PhysRevB.92.100305} due to the competing kinetic term $\propto\Omega$ and on-site energy $\propto \epsilon$. Intuitively, if the kinetic term dominates ($\Omega_{1,2}/(2\epsilon) \gtrsim 1$) strings of excitations expand and merge ballistically, fastly washing local information of the initial configuration, while if it is subleading ($\Omega_{1,2}/(2\epsilon) \lesssim  1$), excitations expand slowly, making possible to retrieve information about the initial conditions.
Such behavior is strongly linked to the localization of the ground state and it can be observed for an exponentially large number, in the system size, of initial states~\cite{PhysRevX.10.021051}. Upon introducing thermal disorder $\tilde{\eta}_\alpha$, the picture is slightly affected. Indeed,
$\tilde{\eta}_\alpha$ can already induce undesired mismatches from the perfect resonant \textit{directional} antiblockade condition \ric{and aid localization}.
Thus, we expect the dynamics to be dictated by the competition of $\Omega$ and the joint contribution of $\tilde{\eta}_\alpha$ and $\epsilon$.
We test this by initializing a representative \ric{high-temperature} product state characterized by regions with low and high densities of excitations.
In the slow phase, heterogeneity in the initial state plays a crucial role in dictating the dynamics up to very long times, in contrast with typical fast thermalizing systems where all local information is quickly lost except for global conserved quantities.
%
%
We \ric{choose} the autocorrelation function~\cite{PhysRevX.10.021051,Garrahan2018}
\begin{equation}
\label{eq_autocorr}
C(t) = \frac{2}{Z}\sum_{i}\langle \ric{\hat{Q}}_i(t)\ric{\hat{Q}}_i(0)\rangle -1,
\end{equation}
where $Z=\sum_i \langle \ric{\hat{Q}}_i(0)\rangle$ is a normalization constant\ric{, as proxy for distinguishing the different dynamical phases}.
For the initial product state considered, $C(t)$ is the density of the initially \ric{excited} sites at time $t$, to which we subtract an evenly distributed `background' corresponding to an infinite temperature state \ric{with} $\langle \ric{\hat{Q}}_i\rangle= 1/2$. Thus, $C(t)$ serves as a good proxy for the memory of initial conditions, as its initial value is $C(t=0)=1$ and tends to zero when the system thermalizes. 
In Fig.~\ref{fig_autocorr_full_theory_canonical_east}(a-b) we show the time-average $\overline{C(t^\star)} = \int_0^{t^\star} C(\tau)d\tau/t^\star$ up to experimentally accessible time windows using the full theory (cf. Eq.~\eqref{eq_H_complete}) in two scenarios: either keeping fixed the thermal disorder and varying $\epsilon$, or vice versa.
As anticipated, both $\epsilon$ and $\tilde{\eta}_x$ contribute \ric{to} slowing down dynamics.
Indeed, in both dynamical phase diagrams, we can distinguish a phase where the system retains memory of the initial state and a phase where the system \ric{quickly} thermalizes and all local memory is quickly erased.     \\


\textit{Perspectives.---}
In this work, we have proposed a scheme for realizing chiral interactions \ric{through} a \textit{directional} antiblockade condition, namely an atom can change its internal state only if the atom to its right (or left) is excited. Our scheme is based on `energetic' arguments and gives rise to constrained interactions. 
Additionally, our protocol could be readily extended in the presence of dominant classical noise. In \ric{this} regime, dynamics is effectively described by rate equations, with rates dependent on the detunings, interactions\ric{,} and atomic configuration~\cite{PhysRevLett.111.215305,PhysRevA.93.040701}, opening up to the simulation of dissipative uni-directional spin dynamics (for further details see~\cite{SupplementalMaterial}).
%
Finally,
we highlight that the degree of chirality in the interactions can be tuned by relaxing the condition $\Delta V \equiv |V_1-V_2| \gg \Omega_{1,2}$. Specifically, for $0 <\Delta V/\Omega_{1,2} \lesssim 1$, dynamics still has a preferable direction,  but there are near-resonant processes towards the other as well (similarly to other Rydberg proposals~\cite{PhysRevX.10.021031,PhysRevA.108.023305}). This offers a path for accessing regimes with a tunable \textit{bias} towards one direction or the other.
As an extreme example, in the zero \textit{bias} case ($\Delta V=0$), and by setting $\Omega_1=\Omega_2/2$, we can effectively simulate the quantum Fredrickson-Andersen model~\cite{Hickey2016,PhysRevLett.53.1244}.

\vspace{10pt}
\textit{Acknowledgements.---}
R.J.V.T. is grateful to O. Chelpanova and especially to M. Stefanini for useful discussions.
This project has
been supported by the Deutsche Forschungsgemeinschaft (DFG, German Research Foundation) through the
Project ID 429529648-TRR 306 QuCoLiMa (``Quantum Cooperativity of Light and Matter''), and the grant
HADEQUAM-MA7003/3-1; by the Dynamics and
Topology Center, funded by the State of Rhineland Palatinate.
This material is based upon work supported by the US Department of Energy, Office of Science, National Quantum Information Science Research Centers.
The work of M. F. has been supported by the DFG (SFB TR 185), project number 277625399.
Parts of this research were conducted using the Mogon supercomputer and/or advisory services offered by
Johannes Gutenberg University Mainz (\url{hpc.uni-mainz.de}), which is a
member of the AHRP (Alliance for High Performance Computing in Rhineland Palatinate,  \url{www.ahrp.info}), and the Gauss
Alliance e.V. We gratefully acknowledge the computing time granted on the Mogon supercomputer at Johannes Gutenberg University Mainz (\url{hpc.uni-mainz.de}) through the project ``DysQCorr.''
%
\bibliography{Rydberg_east_model_v2}

\newpage

\renewcommand{\theequation}{S\arabic{equation}}
\renewcommand{\thefigure}{S\arabic{figure}}
\setcounter{figure}{0}
\setcounter{equation}{0}

\onecolumngrid
\begin{center}
\large{\textbf{SUPPLEMENTAL MATERIAL}}
\end{center}
\section*{Details on the numerical simulations}
Results in Fig.~\ref{figure_chiral_processes}(c-d) \ric{in the main text} were obtained via tensor network methods using the C++ version of the ITensor library~\cite{itensor}. We performed the time evolution via a 3-sites Time Evolving Block Decimation (TEBD) algorithm setting a maximal bond dimension $\chi_\text{max}=500$ (well above what dynamically reached) and timestep $\delta t=10^{-3}$. All the other results \ric{in the main text} were obtained via exact diagonalization using the Python package quimb~\cite{Gray2018}.

\section*{Experimental setup}
Here we provide additional details on how to generate the staggered atomic array and drive fields scheme needed to induce the \textit{directional} antiblockade described in the main text. The atomic array could be produced by trapping the atoms in optical tweezers. Generating these beams using acousto-optic deflectors (AOD) or spatial light modulators (SLM) would allow us to achieve the desired bipartite spacing. An undesired effect due to the trapping potentials holding the atoms is a shift in the atomic energies. Specifically, we can distinguish two components: the intensity variation from trap to trap, and the atomic motion inside the trap, which causes it to sample different intensities/AC Stark shifts in different shots of the experiment. Such effects could mildly compromise the desired \textit{antiblockade} condition. However, unlike other experimental imperfections such as thermal disorder in the atomic positions (which we have taken into account), it is possible to eliminate this shift by switching the traps off entirely during the Rydberg driving~\cite{Bernien2017}. This technique eliminates all potential AC Stark shifts and remains effective for approximately 10 $\mu$s of evolution time, which is the maximum time discussed in our work. The desired staggered configuration in the drive field frequencies could be applied to the atoms through the same microscope objective used for the optical tweezers~\cite{Omran2019}. This, in combination with spatial light modulators (SLM) for the driving beams, would allow the generation of the desired staggered pattern in the drive fields frequencies, i.e., $\Omega_{1,j}(t)=\Omega_1 e^{iV_{j-1,j}t}$, with $\Omega_1$ a constant \ric{and $V_{j-1,j}$ the Van-der-Waals interaction on the $(j-1)$-th bond}. Additionally, the uniform drive field $\Omega_{2}(t) = \Omega_2 e^{i(V_1+V_2)t}$, with $\Omega_2$ as a constant, could also be sent through the microscope objective.

\section*{Alternative scheme for obtaining the directional antiblockade condition}
Here we discuss a different, but equivalent, route for obtaining the \textit{directional} antiblockade condition. Given Eq.~\eqref{eq_H_complete}, we set again $\Omega_2=0$ as it is not necessary for this purpose. In the main text, we discussed a scheme based on a staggered configuration in the atomic distances and drive field frequencies driving the atoms.
Here, we retain the essential ingredient of a staggered configuration in the atomic spacings.
However, instead of imprinting the staggered configuration on the drive field frequencies, which we \ric{now} set to be homogeneous in space, we impose the staggered configuration on the atomic transition frequencies. This condition could be achieved by site-resolved {light shifts~\cite{Chen2023}}.
Specifically, we set the atomic transition frequencies to be detuned by $-V_1$ from the drive field when $j$ is even and by $-V_2$ when $j$ is odd. In this manner, the \textit{directional} antiblockade condition discussed in the main text emerges, as the drive field acting on the $j$-th atom is resonant only when the $(j-1)$-th atom is excited. This can be seen explicitly writing the Hamiltonian in the rotating frame with respect to the drive field, which is given by
\begin{equation}
\label{eq_H_v2}
\hat{H}= -\sum_{j=1}^N V_{j-1,j} \ric{\hat{Q}}_j + \frac{\Omega_1}{2} \sum_{j=1}^N  \ric{\hat{X}_{j}} +V_1 \sum_{j \: \text{odd}} \ric{\hat{Q}}_j \ric{\hat{Q}}_{j+1} + V_2 \sum_{j \: \text{even}} \ric{\hat{Q}}_j \ric{\hat{Q}}_{j+1} + V_\text{NNN}\sum_{j=1}^{N-2} \ric{\hat{Q}}_j \ric{\hat{Q}}_{j+2},
\end{equation}
\ric{where $V_{j,j+1}=V_1$ for $j$ odd and $V_{j,j+1}=V_2$ for $j$ even.}
From Eq.~\eqref{eq_H_v2}, it can be seen that if the $(j-1)$-th is excited, the energy shift cancels the contribution from the first term on the $j$-th atom, making the acting drive field resonant and able to induce a spin-flip.
\section*{Derivation of the effective theory with both sets of drive fields}
Given the Hamiltonian in Eq.~\eqref{eq_H_complete}, we set
\ric{$\Omega_j(t)=\Omega_1 e^{-i (V_{j-1,j}-\epsilon)t} + \Omega_2 e^{-i(V_1+V_2-\epsilon) t}$}\ric{, where $V_{j-1,j}$ is the Van-der-Waals interaction on the $(j-1)$-th bond and $\epsilon$ is a small detuning from the perfect directional antiblockade,}
obtaining
\begin{equation}
\label{eq_H_complete_SS}
\begin{split}
\hat{H}(t)=& \sum_{j=1}^N \left[ \frac{1}{2}\left( \Omega_1 e^{-i(V_{j-1,j}-\epsilon)t} +\Omega_2 e^{-i(V_1+V_2-\epsilon) t}\right) \hat{\sigma}_{j}^+ + \text{H.c.}\right]  +\\
&+V_1 \sum_{j \: \text{odd}} \ric{\hat{Q}}_j \ric{\hat{Q}}_{j+1} + V_2 \sum_{j \: \text{even}} \ric{\hat{Q}}_j \ric{\hat{Q}}_{j+1} + V_\text{NNN}\sum_{j=1}^{N-2} \ric{\hat{Q}}_j \ric{\hat{Q}}_{j+2},
\end{split}
\end{equation}
To obtain the effective Hamiltonian, we pass in the interaction picture	$\hat{H}_\text{int} = \hat{U}^\dagger \hat{H} \hat{U} - \hat{H}_0$ with
\begin{equation}
\hat{U} = \exp\left[-i \hat{H}_0 t \right],\qquad \hat{H}_0 =-\epsilon \sum_{j=1}^N \ric{\hat{Q}}_j + \sum_{j=1}^{N-1} V_{j,j+1} \ric{\hat{Q}}_j \ric{\hat{Q}}_{j+1}.
\end{equation}
We have
\begin{equation}
e^{-i \epsilon \sum_j \ric{\hat{Q}}_j t} \hat{\sigma}_k^+ e^{i\epsilon \sum_j \ric{\hat{Q}}_j t} = e^{-i \epsilon |1\rangle \ric{_k {}_k}\langle 1| t} |1\rangle\ric{_k {}_k}\langle 0| e^{i\epsilon  |1\rangle\ric{_k {}_k} \langle 1| t} = e^{-i\epsilon t} \hat{\sigma}_k^+,
\end{equation}
and, exploiting $\ric{\hat{Q}}_i^n = \ric{\hat{Q}}_i$ for $n\in \mathbb{N}^+$, we have
\begin{equation}
\begin{split}
e^{+i \sum_j V_{j-1,j}\ric{\hat{Q}}_{j-1}\ric{\hat{Q}}_j t} \hat{\sigma}_k^+ e^{-i \sum_j V_{j-1,j}\ric{\hat{Q}}_{j-1}\ric{\hat{Q}}_j t}&= e^{+i V_{k-1,k}\ric{\hat{Q}}_{k-1} t } \hat{\sigma}_k^+  e^{+i V_{k,k+1}\ric{\hat{Q}}_{k+1} t }\\
&=\left[1 + \left(\sum_{n=1}^\infty \frac{(i V_{k-1,k}t)^n}{n!}\right)\ric{\hat{Q}}_{k-1}\right] \hat{\sigma}_k^+ \left[1 + \left(\sum_{n=1}^\infty \frac{(i V_{k,k+1}t)^n}{n!}\right)\ric{\hat{Q}}_{k+1}\right] \\
&=\left[ 1 + \left(e^{+i V_{k-1,k}t}-1\right) \ric{\hat{Q}}_{k-1}\right] \hat{\sigma}_k^+ \left[ 1 + \left(e^{+i V_{k,k+1}t}-1\right) \ric{\hat{Q}}_{k+1}\right] \\
&=\left( \hat{P}_{k-1} + e^{+iV_{k-1,k}t}\ric{\hat{Q}}_{k-1}\right)\hat{\sigma}_k^+  \left( \hat{P}_{k+1} + e^{+iV_{k,k+1}t}\ric{\hat{Q}}_{k+1}\right)
\end{split}
\end{equation}
where we have introduced $\hat{P}_k = (1-\ric{\hat{Q}}_k)$ (projector on the vacuum).
Thus, paying attention to the boundary terms, the Hamiltonian in interaction picture is given by
\begin{equation}
\label{eq_Htilde_rotating_frame}
\begin{split}
\hat{H}_\text{int} &= \frac{1}{2} \left\{ \left(\Omega_1 e^{-iV_2t} +\Omega_2 e^{-i(V_1+V_2) t}\right)
\hat{\sigma}_1^+ \left(\hat{P}_{2} + e^{+iV_1 t}\ric{\hat{Q}}_{2}\right) + H.c.\right\}+\\
&+\frac{1}{2} \sum_{j=2}^{N-1} \left\{ \left(\Omega_1 e^{-iV_{j-1,j}t} +\Omega_2 e^{-i(V_1+V_2) t}\right)
\left( \hat{P}_{j-1} + e^{+iV_{j-1,j}t}\ric{\hat{Q}}_{j-1}\right)\hat{\sigma}_j^+ \left(\hat{P}_{j+1} + e^{+iV_{j,j+1}t}\ric{\hat{Q}}_{j+1}\right) + H.c.\right\}+\\
&+  \frac{1}{2} \left\{ \left(\Omega_1 e^{-iV_{N-1,N}t} +\Omega_2 e^{-i(V_1+V_2) t}\right)
\left( \hat{P}_{N-1} + e^{+iV_{N-1,N}t}\ric{\hat{Q}}_{N-1}\right) \hat{\sigma}_N^+  + H.c.\right\}+\\
&+\epsilon \sum_{j=1}^N \ric{\hat{Q}}_j+V_\text{NNN}\sum_{j=1}^{N-2} \ric{\hat{Q}}_j \ric{\hat{Q}}_{j+2}.
\end{split}
\end{equation}
Given Eq.~\eqref{eq_Htilde_rotating_frame}, \ric{upon setting $V_{j,j+1}=V_1$ for $j$ odd and $V_{j,j+1}=V_2$ for $j$ even,} we obtain the effective theory in Eq.~\eqref{eq_H_eff_complete} after performing the RWA in the limit $|V_1-V_2|,V_{1,2} \gg |\Omega_{1,2}|$, keeping the most resonant processes (the ones without any time dependence).
\section{Effective Hamiltonian for $\Omega_2=0$}
When $\Omega_2=0$, the number of strings of excitations $\ric{\hat{\mathcal{N}}}=\sum_{j} \ric{\hat{Q}}_j(1-\ric{\hat{Q}}_{j+1})$ commutes with the Hamiltonian. Additionally, due to the `East symmetry', the left edge of each string does not move. As a consequence, the dynamics of a $\mathcal{N}$-strings state can be computed from the dynamics of $\mathcal{N}$ 1-string states. Therefore, we can focus on the symmetry sector with a single string ($\ric{\mathcal{N}}=1$). A complete basis of single-kink states is given by $|\bm{k}\rangle = \bigotimes_{j=1}^k|1\rangle \bigotimes_{j=k+1}^N|0\rangle$. Given
\begin{equation}
\begin{split}
\langle \bm{q}| \ric{\hat{Q}}_j \ric{\hat{Q}}_{j+2}|\bm{k}\rangle &= \delta_{q,k}\theta(k-(j+2)),\\
\langle \bm{q}| \ric{\hat{Q}}_j \ric{\hat{X}_{j+1}} |\bm{k}\rangle &= \delta_{q-1,k}\delta_{j-1,k} + \delta_{q+1,k}\delta_{j,k},\\
\langle \bm{q}|\ric{\hat{Q}}_j \ric{\hat{X}_{j+1}}\ric{\hat{Q}}_{j+2}|\bm{k}\rangle &= 0,
\end{split}
\end{equation}
and plugging them in the Hamiltonian in Eq.~\eqref{eq_H_eff_complete} represented in the same basis
\begin{equation}
\hat{H} = \sum_{k,q}\langle \bm{q}|\hat{H}|\bm{k}\rangle |\bm{q}\rangle \langle \bm{k}|,
\end{equation}
we straightforwardly obtain Eq.~\eqref{eq_Heff_integrable}.

\begin{figure}[b!]
\centering
\includegraphics[width=0.48\linewidth]{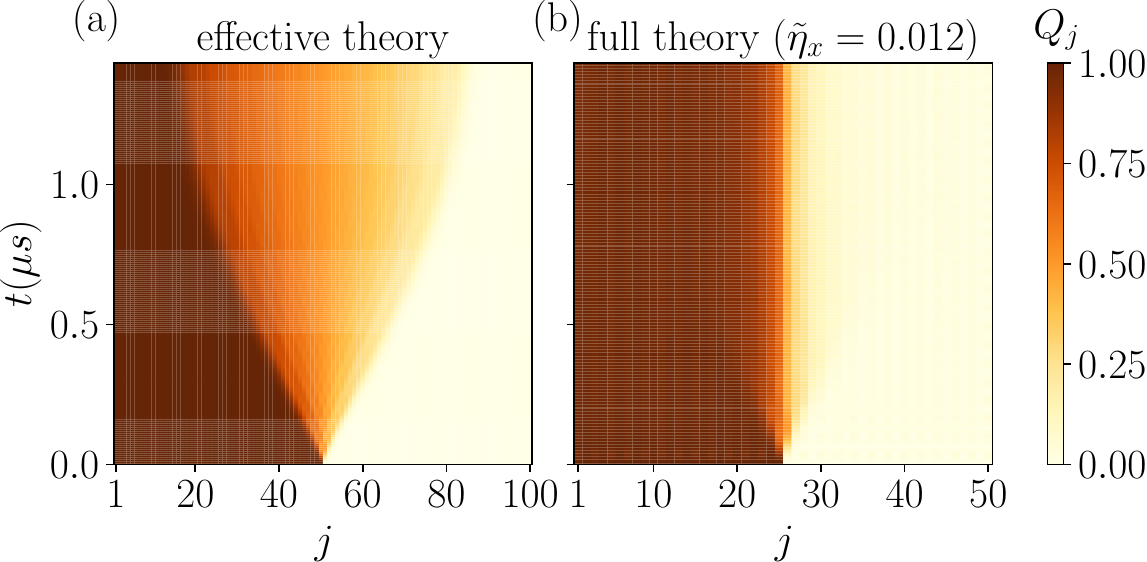}
\includegraphics[width=0.48\linewidth]{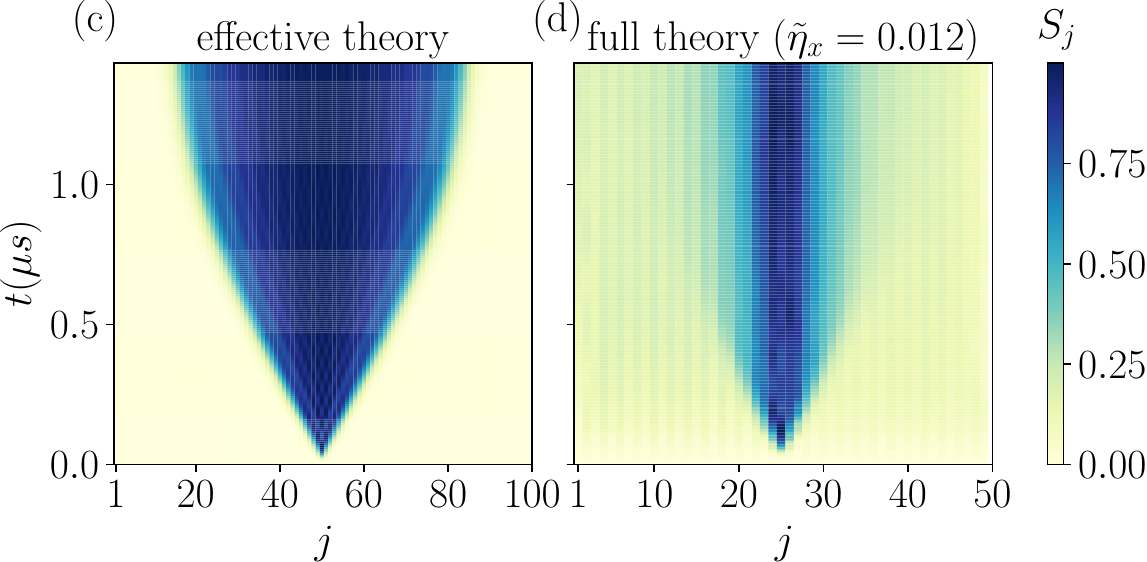}
\caption{{Dynamics in the confined regime ($\Omega_2=0$) of an initial kink state $\bigotimes_{j=1}^M |1\rangle \bigotimes_{j=M+1}^N|0\rangle$ with $M=N/2$ excitations, with $N$ the system size. We compare the dynamics obtained in the ideal regime, where the effective theory (cf. Eq.~\eqref{eq_Heff_integrable}) is exact, with a realistic experimental scenario (cf. Eq.~\eqref{eq_H_complete}) including positional disorder $\eta$ due to thermal fluctuations. The system displays real-time Bloch oscillations along the surface both in \ric{the excitation profile} (a,b), as well as in the entanglement entropy along each cut (c,d). Parameters: $\Omega_1/V_1=0.2$, $V_2/V_1=2$, $\epsilon=0$, $V_1=2\pi \times 15 \:\text{MHz}$.}}
\label{fig_confined_details}
\end{figure}
\section*{Dynamical features in the different regimes}
Here, we further characterize the dynamical features in two distinct extremes in the available parameter space, namely $\Omega_2/\Omega_1=0$ and $\Omega_1/\Omega_2=1$, corresponding to the confined regime and the `quantum East model' regime discussed in the main text. We do not provide further details on the scarred regime $\Omega_1/\Omega_2=0$ as it has been extensively investigated previously~\cite{Turner2018,PhysRevLett.122.220603,Serbyn2021,PhysRevB.98.155134,PhysRevB.98.155134,PhysRevB.99.161101,PhysRevX.11.021021}. Such regimes are characterized by qualitatively different ergodicity-breaking mechanisms, which are also reflected in strikingly different dynamics at the level of experimentally relevant observables, such as the occupation number.\\
For simplicity, we set $\epsilon=0$, which translates to a drive field resonant with the ideal directional antiblockade condition of our proposed implementation. To enhance clarity, we illustrate the dynamics by considering both the effective theory (cf. Eq.~\eqref{eq_H_eff_complete}) and the full theory (cf. Eq.~\eqref{eq_H_complete}) with quenched disorder, mimicking with the latter a realistic experimental scenario. We use the same parameters as those used in the main text.
\subsection*{Confined regime ($\Omega_2/\Omega_1=0$)}
In this regime, strings of excitations can only expand or shrink, making the dynamics occur solely along the surface. Thus, to better observe such dynamical features, we initialize a kink state $\bigotimes_{j=1}^M |1\rangle \bigotimes_{j=M+1}^N|0\rangle$ with $M=N/2$ excitations, where $N$ is the system size, set to be $N \geq 50$ (see Fig.~\ref{fig_confined_details}).
Simulations of the effective theory (cf. Eq.~\eqref{eq_Heff_integrable}) can be efficiently carried out using Exact Diagonalization. Instead, we employ Tensor Network methods for simulating the full theory (cf. Eq.~\eqref{eq_H_complete}). As stated in the main text, the system displays confinement due to emerging real-time Bloch oscillations with a period $T= 2\pi/V_\text{NNN}$ and width $\ell \sim \Omega_1/V_\text{NNN}$. In the full theory, such oscillations are still present, although suppressed due to quenched disorder in the atomic positions.\\
For completeness, we also show the dynamics of the entanglement entropy along each cut to demonstrate how quantum correlation builds up solely along the surface and remains confined, impeding thermalization.

\begin{figure}[h!]
\centering
\includegraphics[width=0.48\linewidth]{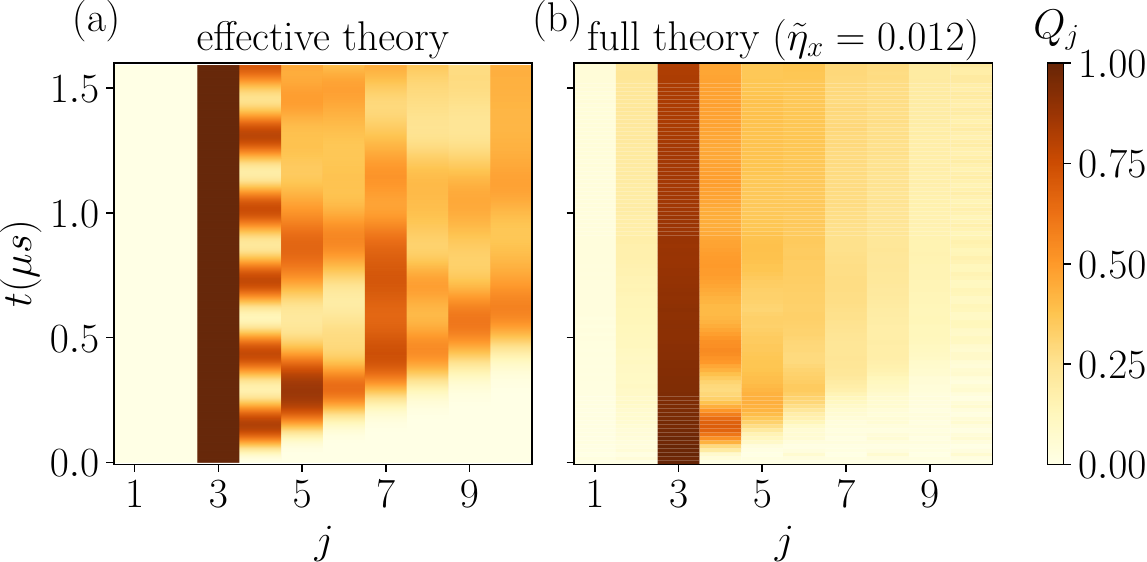}
\includegraphics[width=0.28\linewidth]{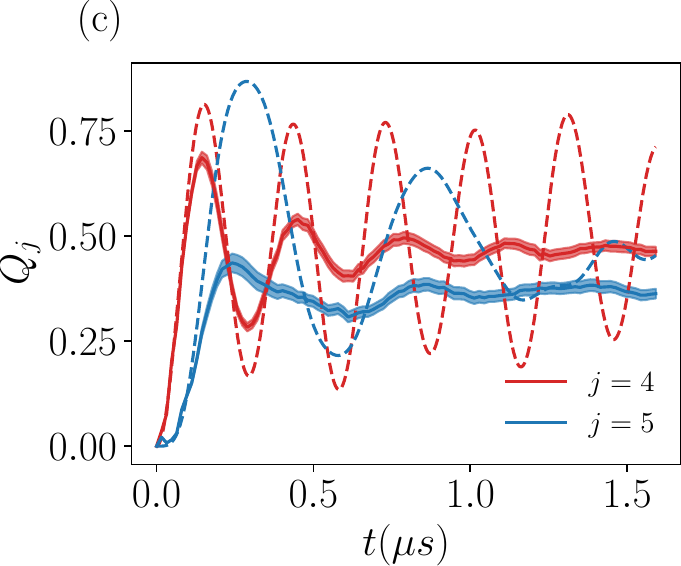}
\caption{{Dynamics in the `quantum East model' regime ($\Omega_1/\Omega_2=1$) of an initial seeded excitations We compare the dynamics obtained in the ideal regime, where the effective theory (cf. Eq.~\eqref{eq_H_eff_complete}) is exact, with a realistic experimental scenario (cf. Eq.~\eqref{eq_H_complete}) including positional disorder $\eta$ due to thermal fluctuations. The system displays both a propagating front of excitations towards the East as well as oscillations in the bulk (a,b). In (c) we show some vertical cuts of the heatmaps (a,b) for clarity, comparing the effective theory (dashed line) with the full theory (continuous line) upon averaging $100$ disorder realizations. The shaded areas around the continuous lines are the statistical errors due to the finite number of sampled trajectories. Parameters: $\Omega_{1,2}/V_1=0.2$, $V_2/V_1=2$, $\epsilon=0$, $V_1=2\pi \times 15 \:\text{MHz}$.}}
\label{fig_quantum_east_model_details}
\end{figure}

{
\subsection*{`Quantum East Model' ($\Omega_1/\Omega_2 \neq 0$)}
In this regime, dynamics occur both in the bulk and along the surface. Thus, we consider an initial state with a single excitation (cf. Fig.~\ref{fig_quantum_east_model_details}). We observe, similar to the confined regime, a propagating front of excitations towards the East, with additional oscillations in the bulk, attributed to the presence of `string-breaking terms.' The oscillations are influenced by the strength of the external drive field, exhibiting a period $\sim \Omega_{2}$ (which is the Rabi frequency of the global drive controlling the `string-breaking' processes) and amplitudes of near order $1$.}

\section*{Classical rate equations}
\begin{figure}[h!]
\centering
\includegraphics[width=0.7\linewidth]{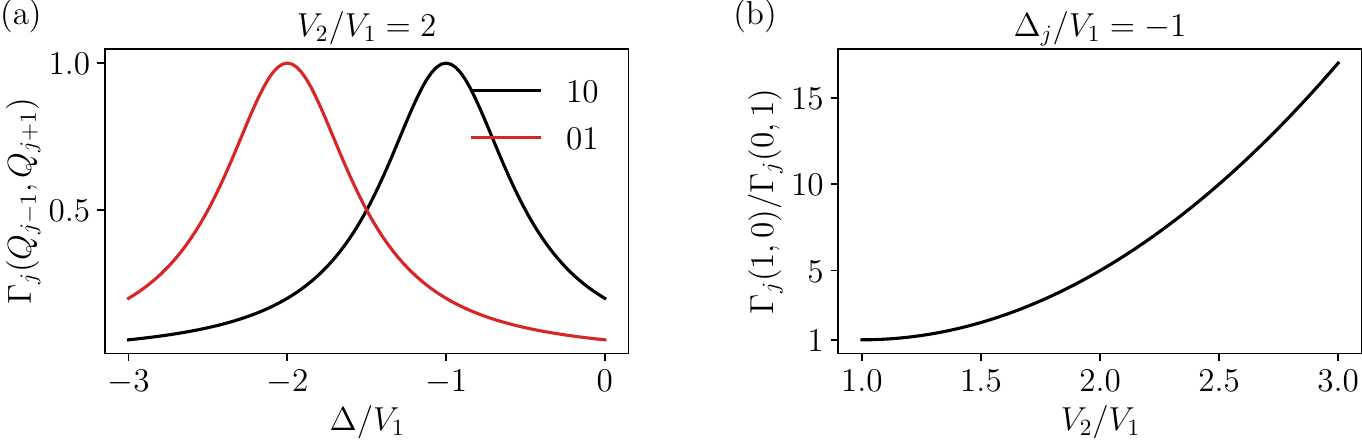}
\caption{(a) Rate $\Gamma_j$ (cf. Eq.~\eqref{eq_rate_j_specific}) at fixed $V_2/V_1=2$ and atomic configurations (values of $\ric{Q}_{j-1}$ and $\ric{Q}_{j+1}$). As evident, the rate is not symmetric under the exchange of neighboring atoms (the two curves do not overlap). (b) Ratio of the rates $\Gamma_j$ in the two scenario $\ric{Q}_{j-1}=1$, $\ric{Q}_{j+1}=0$  and viceversa at fixed $\Delta_j=-V_1$. In the symmetric case ($V_1=V_2$) no notion of directionality appears ($\Gamma_j(1,0)/\Gamma_j(0,1)=1$). Instead, for $V_1 \neq V_2$, rates display `chiral' features ($\Gamma_j(1,0)/\Gamma_j(0,1) \neq 1$). The `chirality' can be readily tuned by simply changing the ratio $V_1/V_2$, or the detuning $\Delta_j$.  In all the calculations we have set $\gamma/V_1=1$.}
\label{fig_sup_rates}
\end{figure}

Let us restrict for simplicity to the single drive field regime setting $\Omega_2=0$. We keep the detuning $\Delta_j$ of the drive field frequency from the bare atomic transition as a free parameter for generality. In the rotating frame  with respect to the drive fields, the starting Hamiltonian is given by
\begin{equation}
\label{eq_H_v3}
\hat{H}= \sum_{j=1}^N \Delta_j \ric{\hat{Q}}_j + \frac{\Omega}{2} \sum_{j=1}^N  \ric{\hat{X}_{j}} + \frac{1}{2}\sum_{i,j}V_{i,j}\ric{\hat{Q}}_i \ric{\hat{Q}}_{j}.
\end{equation}
Due to the presence of dephasing, the full dynamics of the state $\hat{\rho}$ is given by the Lindblad master equation
\begin{equation}
\partial_t \hat{\rho} = -i[\hat{H},\hat{\rho}] + \gamma \sum_j \left(\ric{\hat{Q}}_j \hat{\rho}\ric{\hat{Q}}_j - \frac{1}{2}\{\ric{\hat{Q}}_j,\hat{\rho}\}\right)
\end{equation}
In the main text, we have worked in the regime $\gamma \ll \Omega$. Now, we work in the opposite regime $\gamma \gg \Omega$. In this regime, we can distinguish fast and slow processes: the fast processes are controlled by the interacting part of the Hamiltonian and the dissipative process; the slow processes are controlled by the external drive fields. Projecting out the fast dynamics (see Ref.~\cite{PhysRevLett.111.215305} for the derivation), the dynamics of the projected state $\hat{\mu}$ is governed by
\begin{equation}
\begin{split}
\partial_t \hat{\mu} =\frac{4\Omega^2}{\gamma}\sum_j \Gamma_j\left(\ric{\hat{X}_j} \hat{\mu}\ric{\hat{X}_j} - \hat{\mu}\right),
\end{split}
\end{equation}
with rates dependent on the parameters of the Hamiltonian, dissipation and atomic configuration
\begin{equation}
\label{eq_rate_j}
\Gamma_j^{-1} = 1 + 4\left(\frac{\Delta_j + \sum_{m \neq j} V_{j,m}\ric{\hat{Q}}_m}{\gamma}\right)^2.
\end{equation}
Let us focus on a specific site $j$ and fix the staggered configuration in the potentials discussed in the main text. Keeping up to nearest-neighbour terms, the rate in Eq.~\eqref{eq_rate_j} turns into
\begin{equation}
\label{eq_rate_j_specific}
\Gamma_j^{-1} = 1 + 4\left(\frac{\Delta_j + V_{1}\ric{\hat{Q}}_{j-1}+V_2 \ric{\hat{Q}}_{j+1}}{\gamma}\right)^2,
\end{equation}
where we have \ric{set} $V_{j,j+1}=V_1$ \ric{for $j$ odd}  and $V_{j,j+1}=V_2$ \ric{for $j$ even}. As evident, the rate $\Gamma_j$ depends on the atomic configuration and $\Delta_j$. We aim to show that the rates are not symmetric under the exchange of the pair of excitations $(j - 1)\leftrightarrow (j+1)$ \ric{when $V_1 \neq V_2$}. To do so, in Fig.~\ref{fig_sup_rates}(a) we show the rate as a function of $\Delta_j$ for $V_2=2 \times V_1$ in the non-symmetric scenarios, namely
$\ric{Q}_{j-1}=1$, $\ric{Q}_{j+1}=0$  and viceversa. It is evident that at fixed parameters, the rates are generally not symmetric as desired for $V_1 \neq V_2$ (except for a fine-tuned value of $\Delta_j$ at which the curves cross). As in the coherent case, the degree of directionality can be tuned via $|V_1-V_2|$ and $\Delta$ (cf. Fig.~\ref{fig_sup_rates}(b)).

\end{document}